\documentclass[twocolumn,prl,showpacs,nofootinbib,superscriptaddress]{revtex4}
\usepackage{latexsym}
\usepackage{dcolumn}
\RequirePackage{graphicx}
\usepackage{epsfig}
\newcommand{\ba}{\begin{eqnarray}}

\newcommand{\ea}{\end{eqnarray}}
\newcommand{\be}{\begin{equation}}
\newcommand{\ee}{\end{equation}}

\newcommand{\eq}[1]{Eq.\,(\ref{#1})}

\def\bea{\begin{eqnarray}} 
\def\eea{\end{eqnarray}} 
\newcommand{\la}{\,\raisebox{-.8ex}{\,$\stackrel{\textstyle <}{\sim}$}\,\,} 
\newcommand{\ga}{\,\raisebox{-.8ex}{\,$\stackrel{\textstyle >}{\sim}$}\,\,} 

\begin{document}

\title{Analytic derivation of the leading-order gluon distribution function $G(x,Q^2)=xg(x,Q^2)$ from the proton structure function $F_2^p(x,Q^2)$}

\author{Martin~M.~Block}
\affiliation{Department of Physics and Astronomy, Northwestern University, 
Evanston, IL 60208}
\author{Loyal Durand}
\affiliation{Department of Physics, University of Wisconsin, Madison, WI 53706}
\author{Douglas W. McKay}
\affiliation{Department of Physics and Astronomy, University of Kansas, Lawrence, KS 66045} 
\date{\today}

\begin{abstract}
We derive a second-order linear differential equation for the leading-order gluon distribution function $G(x,Q^2)=xg(x,Q^2)$ which determines $G(x,Q^2)$ directly from the proton structure function $F_2^p(x,Q^2)$.  This equation is derived from the leading order DGLAP evolution equation for $F_2^p(x,Q^2)$, and does not require knowledge of either the individual quark distributions or the gluon evolution equation. Given an analytic expression that successfully reproduces the known  experimental data for $F_2^p(x,Q^2)$ in a domain $x_{\rm min}(Q^2) \le x \le x_{\rm max}(Q^2)$, $Q^2_{\rm min}\le Q^2\le  Q^2_{\rm max}$ of the Bjorken variable $x$  and the virtuality $Q^2$ in deep inelastic scattering, $G(x,Q^2)$ is uniquely determined in the same domain.  We give the general solution and illustrate the method  using the recently proposed  Froissart bound type parametrization of $F_2^p(x,Q^2)$ of E. L. Berger, M. M. Block and C-I Tan, PRL {\bf 98}, 242001, (2007). Existing leading-order gluon distributions based on power-law descriptions of individual parton distributions agree roughly with the new distributions for $x\ga 10^{-3}$ as they should, but are  much larger  for $x\la 10^{-3}$. 
\end{abstract}


\maketitle

%%%%%%%%%%%%%%%%%%%%

{\em Introduction.} \ \ Parton distributions  play a key role in our understanding of Standard Model processes, in our predictions for Standard Model processes at accelerators, and in our searches for new physics.  In particular, accurate knowledge of gluon distribution functions at small Bjorken $x$ will play a vital role in estimating backgrounds, and hence, our ability to search for new physics at the Large Hadron Collider.  

Traditionally, gluon and quark distribution functions have been determined  simultaneously by starting with a virtuality $Q^2_0$, typically in the 1 to 2 GeV$^2$ range, using the two coupled integral-differential DGLAP equations  \cite{dglap} to  evolve  individual quark and gluon trial distributions to higher $Q^2$. The results are adjusted to fit  the overall data (mainly the experimental data for the  proton structure function $F_2^p(x,Q^2)$) by adjusting the parameters in the initial parton distributions, thus determining the evolved distributions.  

In this paper, we present a new and simple method for determining the gluon distribution function $G(x,Q^2)=xg(x,Q^2)$ in leading order (LO) directly from a global parametrization of the data on $F_2^p(x,Q^2)$.  The method  {\em neither requires knowledge of the separate quark distributions, nor the use of the DGLAP evolution equation for $G(x,Q^2)$}, in the region in which proton structure function data exist.   We illustrate the method using a Froissart bound-type fit \cite{bbt} to the proton structure function, and compare with other LO results. 

Results on $G(x,Q^2)$ obtained this way do not obviate the need for simultaneous fits to all quark and gluon distributions which may use data other than that on $F_2^p(x,Q^2)$, but should provide a useful check on those results.

%%%%%%%%%%%%%%%%%%%%%%%

{\em Strategy for determining gluon distributions.}\ \ Instead of starting with parametrizations of parton distribution functions at some $Q_0^2$ and then evolving from $Q_0^2$ to the desired $Q^2$, we follow a new strategy as follows: 
\begin{enumerate}
\item We first make a {\em global} parametrization of the experimental proton structure function $F_2^p$ {\em  simultaneously} in $x$ and $Q^2$.  Berger, Block and Tan \cite{bbt} give an example of such such a parametrization which gives an excellent fit to all of the available ZEUS data \cite{ZEUS} in the domain $10^{-4}\la x \le .09$, $0.11\le Q^2\le 1200$ GeV$^2$, and also to H1 results \cite{H1} that were not used in the fit.  A smooth, accurate fit allows us to calculate the derivatives of $F_2^p(x,Q^2)$ we will need below.
\item We next derive  an inhomogeneous second-order linear differential equation for $G(x,Q^2)$ from the LO DGLAP equation that determines the evolution of the (now known) proton structure function $F_2^p(x,Q^2)$.  The inhomogeneous driving term in the equation is determined entirely by $F_2^p(x,Q^2)$ and its derivatives.
\item  Finally, we solve the differential equation explicitly for the gluon distribution function. 
\end{enumerate}
This approach requires {\em no  assumptions} about either the shape of the {\em gluon distribution} or  the shape of {\em individual} quark distributions.  The method gives a global LO gluon distribution function $G(x,Q^2)$ which is  unique---within experimental uncertainties---in the region in $x$ and $Q^2$ space  which contains the  experimental data on $F_2^p(x,Q^2)$.

%%%%%%%%%%%%%%%%%%%%

 {\em Differential equation for the leading order gluon distribution function.}\ 
The  LO DGLAP equation\cite{dglap} for the evolution of the proton structure function $F_2^p(x,Q^2)$ can be written  as
\ba
\frac{\partial F_2^p(x,Q^2)}{\partial \ln (Q^2)}&=&x\left[\int_{x}^1\frac{ F_2^p(z,Q^2)\times K_{qq}({x}/z)}{z^2}\,dz\right.\nonumber\\
&&\hspace*{-6em}+\left. \sum_{i}\int_{x_i}^1 e_i^2(z,Q^2)\frac{G(z,Q^2)\times K_{qg}({x_i}/z)}{z^2}\,dz\right]\!\!.\ \ 
\label{AP2}
\ea
Here $G(x,Q^2)=xg(x,Q^2)$ and  $e_i^2(z,Q^2)=e_i^2\theta(z-x_i)$ where $e_i$ is the electric charge of the quark with flavor $i$,  $x_i=x[1+(4M_i^2/Q^2)]$, and the step functions enforce the parton level threshold conditions $s\geq 4M_i^2$ for the production of a pair of quarks of mass $M_i$ \cite{tung}.  The sum runs over all quarks and antiquarks.  $K_{qq}$ and $K_{qg}$ are the LO splitting functions for quarks and gluons, respectively.  

To illustrate our method most simply, we will here ignore the threshold factors and  suppose that we have just four active quarks, $u,d,c,s$, with all quarks treated as massless.  Then $\sum_{i}e_i^2=20/9$.  

We will show elsewhere that the method can be generalized within the {\em identical framework\ } to include the mass-dependent threshold effects---again without requiring knowledge of the individual quark distributions---with minimal complication.

With this simplified starting point, we write \eq{AP2} as
\ba
\frac{\partial F_2^p(x,Q^2)}{\partial \ln (Q^2)}\!\!\!&=&\!\!\!\frac{\alpha_s}{4\pi}\left\{ 4{F_2^p(x,Q^2)}\!+\!{16\over 3}\left[{F_2^p(x,Q^2)}\ln\frac{1-x}{x}\right.\right.\nonumber\\
&&\left.\left.
\hspace*{-4em}+x\int_x^1\left(\frac{F_2^p(z,Q^2)}{z}-\frac{F_2^p(x,Q^2)}{x}\right){dz\over z-x}\right]\right.\nonumber\\
&&\hspace*{-4em}\left.-\frac{8}{3}x\int_x^1F_2^p(z,Q^2)\left(1+\frac{x}{z}\right)\frac{\,dz}{z^2}\right.\nonumber\\
&&\hspace*{-4em}\left.+\frac{20}{9}x\int_x^1G(z,Q^2)\left(\frac{x^2+(z-x)^2}{z^2}\right)\frac{\,dz}{z^2}\right\},\label{AP200}
\ea
following Durand and Putikka \cite{randy} in the first of their Eqs. (19).\footnote{ The  term $-\frac{1}{6} G(x,Q^2)$ in the first of Eqs.\ (19) in \cite{randy} should be  deleted there and inserted in the second of Eqs. {19} with a coefficient $2n_f$. A term $-6x\int_x^1dx_1G(x_1,t)/x_1^2)$ was also  dropped in the transition from Eq. (18) to the second of Eqs.\ (19) and should be inserted.  Note that in Ref. \cite{randy},  the notation $G(x,Q^2)$ was used for the function $g(x, Q^2)$ of the present paper.}

Integrating the term in line 3 of \eq{AP200} by parts and using the boundary condition $F_2^p(1,Q^2)=0$, we find that 
\ba
-\frac{8}{3}x\int_x^1F_2^p(z,Q^2)\left(1+\frac{x}{z}\right)\frac{\,dz}{z^2}&=&-4 F_2^p(x,Q^2)\nonumber\\
&&\hspace*{-12em}-{4\over3}\int_x^1\frac{\partial F_2^p}{\partial z}(z,Q^2)\left(\frac{x^2}{z^2}+
\frac{2x}{z}\right)\,dz.\label{AP20}
\ea
We rewrite the large bracket in \eq{AP200} as 
\ba
&&F_2^p(x,Q^2)\ln(1-x)-{F_2^p(x,Q^2)}\ln x
\nonumber\\
&&\qquad+x\int_x^1F_2^p(z,Q^2){dz\over z(z-x)}\nonumber\\
&& \qquad-F_2^p(x,Q^2)\left(\int_x^1\frac{dz}{z}+ \!\! \int_x^1\frac{x}{z(z-x)}dz   \right) \label{partialfraction2}\\
&&=F_2^p(x,Q^2)\ln(1-x) \nonumber \\
&&\qquad+x\int_x^1\frac{F_2^p(z,Q^2)-F_2^p(x,Q^2)}{z(z-x)}\,dz  \nonumber\\
&&=\int_x^1\frac{ \partial F_2^p}{\partial z}(z,Q^2)\ln\frac{z}{z-x}\,dz,\label{bracket}%eq9
\ea
where we used the identity ${1\over z-x}={1\over z}+{x\over z(z-x)}$ to get from the large bracket in \eq{AP200} to \eq{partialfraction2}, and partial integration, with $F_2^p(1,Q^2)=0$, to get to \eq{bracket}.

Next, we combine terms and define ${\cal F}_2^p (x,Q^2)$ as 
\ba
{\cal F}_2^p(x,Q^2)&\equiv&\frac{\partial F_2^p(x,Q^2)}{ \partial\ln (Q^2)}\nonumber\\
&&\hspace*{-2ex}-\frac{\alpha_s}{4\pi}\left\{ {16
\over 3} \int_x^1\frac{
 \partial F_2^p}{\partial z}(z,Q^2)\ln\frac{z}{z-x}\,dz \right.\ \ \nonumber\\
&& \hspace*{-2ex}§\left.-{4\over3}\int_x^1\frac{\partial F_2^p}{\partial z}(z,Q^2)\left(\frac{x^2}{z^2}+
\frac{2x}{z}\right)\,dz\right\}\!.\label{calF}
\ea
After multiplying both sides by $4\pi/\alpha_s$, we  rewrite \eq{AP200} as 
\ba
\frac{4\pi}{\alpha_s}{\cal F}_2^p(x,Q^2)/x=\hphantom{xxxxxxxxxxxxxxxxxxxxxx}&&\nonumber\\
\frac{20}{9}\int_x^1G(z,Q^2)\left(\frac{x^2+(z-x)^2}{z^2}\right)\frac{\,dz}{z^2}.\hphantom{xxxxxx}&&\label{finalAPoverx}
\ea

After successive differentiations of both sides of \eq{finalAPoverx} with respect to $x$,  multiplication by $-x^4$, and some rearranging, we find an inhomogeneous second-order differential equation which determines $G(x,Q^2)$ in terms of ${\cal F}^p_2(x,Q^2)$:
\ba
x^2\frac{\partial^2}{\partial x^2}G(x,Q^2)-2x\frac{\partial}{\partial x}G(x,Q^2)+4G(x,Q^2) &&\nonumber \\
=-\frac{4\pi}{\alpha_s}\frac{9}{20}x^4\frac{\partial^3}{\partial x^3}\left(\frac{{\cal F}_2^p(x,Q^2)}{x}\right). \label{Geq}
\ea
To simplify the notation in this and other equations,  we define  ${\cal G}_n(x,Q^2)$ for $n$ active quarks by  
\be
{\cal G}_n(x,Q^2)\equiv
-\frac{4\pi}{\alpha_s} x^4\frac{\partial^3}{\partial x^3}\left(\frac{{\cal F}_2^p(x,Q^2)}{x}\right)\Big/\sum_{i=1}^n e_i^2.  \label{definecalG_n}
\ee
Again, the sum is over all active quarks and antiquarks. The right hand side of the Eq.\ (\ref{Geq}) is then ${\cal G}_4(x,Q^2)$, where $\sum_{i=1}^4 e_1^2={20/9}$. 

Explicit evaluation of  this term gives
\ba
{\cal G}_4(x,Q^2)=\hspace*{17em}&&
\nonumber \\
-\frac{4\pi}{\alpha_s}\frac{9}{20}x^4\frac{\partial^4 (F_2^p(x,Q^2)/x)}{\partial x^3\partial\ln Q^2}+\frac{12}{5}x\frac{\partial F_2^p}{\partial x}(x,Q^2) \hspace*{1em} &&
\nonumber \\
-3x^2\frac{\partial^2 F_2^P}{\partial x^2}(x,Q^2)+\frac{9}{5}x^3\frac{\partial^3 F_2^p}{\partial x^3}(x,Q^2) \hspace*{1em} && \nonumber \\
+\frac{12}{5}x^4\frac{\partial^3}{\partial x^3}\frac{1}{x}\int_x^1\frac{\partial F_2^p}{\partial z}(z,Q^2)\ln\frac{z}{ z-x}\,dz. \hspace*{1em} \label{expandedd3calF}
\ea

The derivatives of the factor  $\ln[z/(z-x)]$ in the integrand in the last term can be treated using the identity 
\ba
\lim_{\epsilon\rightarrow 0}\frac{\partial}{\partial x}\int_{x+\epsilon}^1 dz\ln(z-x)f(z,Q^2) = \hspace*{6em} && \nonumber \\
\label{identity}
-\ln(1-x) f(1,Q^2) + \int_x^1dz \ln(z-x)\frac{\partial f}{\partial z}(z,Q^2).&&
\ea
In the present case,  $F_2^p(x,Q^2)$ and its first two derivatives with respect to $x$ are expected vanish at $x=1$. The final integrand has only a logarithmic singularity.

%%%%%%%%%%%%%%%%%%%%
  
{\em Analytic solution for $G(x, Q^2)$.}\ \ With the notation above, \eq{Geq} becomes the inhomogeneous  differential equation
\ba
x^2\frac{\partial^2}{\partial x^2}G(x,Q^2)-2x\frac{\partial}{\partial x}G(x,Q^2)+4G(x,Q^2)&& \nonumber \\
 = {\cal G}_4(x,Q^2). && \label{Gdiffeqn}
\ea
Introducing the new variable $v=\ln(1/x)$,  we rewrite  \eq{Gdiffeqn},  a linear 2$^{\rm nd}$ order inhomogeneous equation, as
\be
\left(\frac{\partial^2 }{\partial v^2}+3\frac{\partial}{\partial v}+4\right)\hat{G}(v,Q^2)=\hat{\cal G}_4(v,Q^2), \label{Gofv}
\ee 
with $\hat {G}(v,Q^2)\equiv G(e^{-v},Q^2)$, $\hat{\cal G}_4(v,Q^2)\equiv {\cal G}_4(e^{-v},Q^2)$.
Defining the operator $D\equiv \partial/\partial v$, we factor \eq{Gofv} as
\be
(D-\lambda_+)(D-\lambda_-)\hat {G}(v,Q^2)=\hat{\cal G}_4(v,Q^2) \label{Geqninv}
\ee
where $\lambda_\pm=k\pm i\omega$ with $k=-3/2$, $\omega=\sqrt{7}/2$.

To construct the solution of  \eq{Geqninv}, we introduce the  solutions $e^{\lambda_\pm v}$ of the homogeneous equation and the functions
\be
\label{calGdefined}
{\cal G}_\pm(v,Q^2) = \int_0^vdv'e^{-\lambda_\pm v'}\hat{\cal G}_4(v',Q^2).
\ee
The  solution that satisfies the boundary conditions that  $\hat{ G}(v,Q^2)$ and $\partial \hat{G}(v,Q^2)/\partial v$ vanish at $v=0$, or equivalently, that $G(x,Q^2)$ and    $\partial G(x,Q^2)/\partial x$ vanish at $x=1$, is then
\ba
\label{Gsoln5}
\hat{G}(v,Q^2) &=& \frac{1}{\lambda_+-\lambda_-}\left[e^{\lambda_+v}{\cal G}_+(v,Q^2)  - e^{\lambda_-v}{\cal G}_-(v,Q^2)\right]\nonumber\\
&&\hspace*{-10ex}=\frac{1}{\omega}\int_0^v e^{k(v-v')}\sin\left(\omega(v-v')\right)\hat{\cal G}_4(v',Q^2)\,dv'. \label{vanswer}
\ea
This result is completely general and gives the {\em exact} LO expression for $\hat{G}(v,Q^2)$ once $\hat{\cal G}_4(v,Q^2)$ is known. 

We re-emphasize at this point that both $\hat{\cal G}_4(v,Q^2)$ and $\hat{G}(v,Q^2)$ are completely determined by $F_2^p(x,Q^2)$ through the expressions in Eqs.\  (\ref{expandedd3calF}) and (\ref{vanswer}), and the definition $\hat{\cal G}_4(v,Q^2)={\cal G}_4(e^{-v},Q^2)$.  Given a smooth analytic parametrization of $F_2^p(x,Q^2)$ as a simultaneous function of $x$ and $Q^2$ in a domain $x_{\rm min}(Q^2)\leq x\leq x_{\rm max}(Q^2),\  Q^2_{\rm min}\leq Q^2\leq Q^2_{\rm max}$, we can calculate  
$\hat{\cal G}_4(v,Q^2)$ and thus determine the gluon distribution function $G(x,Q^2)$ to within the accuracy of the parametrization. Different smooth parametrizations   that fit the data on $F_2^p(x,Q^2)$ equally well in the specified domain should give equivalent  gluon distributions within that domain.  

%We stress, furthermore, that it is {\em not} necessary to use the DGLAP evolution equation for $G(x,Q^2)$ itself within the experimental domain:  the evolution is already built into the {\em measured} $F_2^p(x,Q^2)$. It is $F_2^p(x,Q^2)$ which determines ${\cal G}_4(x,Q^2)$, and therefore determines the LO approximation to $G(x,Q^2)$ through the {\em exact}  relation in Eq.\ (\ref{vanswer}). Higher-order (NLO) corrections to $G(x,Q^2)$ will be discussed separately. 
We stress that the exact LO equation, (16), directly builds in the $Q^2$ evolution for $ G(x,Q^2)$ in terms of the measured $F_{2}^{p}(x,Q^2)$ in the experimental domain.  Since higher order (NLO) and higher twist effects are embodied in $F_{2}^{p}(x,Q^2)$, there is no reason to to expect that LO DGLAP evolution of $G(x,Q^2)$ at fixed x should agree with our evaluation of the $Q^2$ dependence.  In fact, the difference is an indication  of corrections to the LO DGLAP evolution in this domain. Explicit NLO corrections to relation (16), and therefore to $G(x,Q^2)$, will be discussed separately. 
One might use the  gluon evolution DGLAP equation,  with its dependence on the singlet quark distribution, if one wished to extend the gluon distribution determined above beyond the experimental domain in $Q^2$ at fixed $x$. 

 We note in addition that the singlet quark distribution can itself be related within the experimental domain to the now-known gluon distribution, using the evolution equation for $G(x,Q^2)$ and a construction similar to that used to relate $G(x,Q^2)$ to $F_2^p(x,Q^2)$, but have so far not investigated this procedure in detail.

%%%%%%%%%%%%%%%%%%%%%%   

{\em Example: solution for $G(x, Q^2)$ using a Froissart bounded structure function $F_2^p(x,Q^2)$.}\ \ We illustrate the procedure above using the Froissart-bound type parametrization of the proton structure function given by Berger, Block, and Tan {\cite{bbt} for $x\le x _P=0.09$, 
\ba
F_2^p(x,Q^2)=(1-x)\left[{F_P\over 1-x_P}\right.\hphantom{xxxxxxxxxxxxxxxx}\nonumber\\
+\left(a_0+a_1\ln(Q^2)+a_2\ln^2(Q^2)\right)\ln \left[\frac{x_P(1-x)}{x(1-x_P)}\right]\hphantom{xx}&&\nonumber\\
\label{bbtfit}
+\left(b_0+b_1\ln(Q^2)+b_2\ln^2(Q^2)\right)\ln^2 \left[\frac{x_P(1-x)}{x(1-x_P)}\right],\label{F2}\hphantom{x}&&
\ea
where $F_P=0.41$ is the value of $F_2^p$ at the approximate fixed point in $Q^2$ at $x=x_P$, and
\ba
a_0&=&-5.381\times 10^{-2}\pm  2.17\times 10^{-3}  ,\nonumber\\
a_1&=&2.034\times 10^{-2}\pm  1.19\times 10^{-3}  ,\nonumber\\
a_2&=&4.999\times 10^{-4}\pm  2.23\times 10^{-4}  ,\nonumber\\
b_0&=& 9.955\times 10^{-3}\pm  3.09\times 10^{-4}   ,\nonumber\\
b_1&=&   3.810\times 10^{-3}\pm  1.73\times 10^{-4}  ,\nonumber\\
b_2&=&  9.923\times 10^{-4}\pm  2.85\times 10^{-5}  ,\label{parameters}
\ea

In the absence so far of a global fit to $F_2^p(x,Q^2)$ for $x_P<x\le 1$, we follow the work of Berger, Block, McKay and Tan \cite{bbmt} and approximate $F_2^P(x,Q^2)$ in that region as 
 \be
  F_2^p(x,Q^2)\approx F_P\left(\frac{x}{x_P}\right)^{\rho(Q^2)}\left(\frac{1-x\;}{1-x_P}\right)^3,
\ee 
or by that form multiplied by an extra factor $\left[1+b(Q^2)(x-x_P)^2\left((1-x)/(1-x_P)\right)^c\right]$.  The exponent
 $\rho(Q^2)$  is  chosen so that the values of the functions and their first derivatives match at $x=x_P$. In the extended form, the coefficient $b(Q^2)$ is used to match the second derivatives at $x=x_P$, and $c$ is used to obtain a rough fit to the high-$x$ data. We have also considered other parametrizations.  The results for ${\cal G}_4(x,Q^2)$ and $G(x,Q^2)$ turn out to be insensitive to this parametrization except for $x$ near $x_P$, where $G(x,Q^2)$ is already very small, and are essentially determined for $x<x_P$ (or $v>v_P$) by the  experimentally determined expression for $F_2^p(x,Q^2)$ in Eq.\ (\ref{F2}).

We have found that we can parametrize the function
$\hat{\cal G}_4(v,Q^2)$ calculated this way to high numerical accuracy for $v>v_P$ as a second degree polynomial in $v=\ln(1/x)$ whose coefficients are  quadratic polynomials in $\ln Q^2$, i.e., as
\be
\hat{\cal G}_4(v,Q^2)=\alpha (Q^2)+\beta (Q^2)v+\gamma(Q^2)v^2, \label{G0ofv}
\ee
 with an appropriate smooth extension to $v<v_P$.  

It is also possible to skip the separate evaluation of $\hat{\cal G}_4(v,Q^2)$ by integrating repeatedly by parts in the expression for $\hat{G}(v,Q^2)$ in Eq.\ (\ref{vanswer}) to eliminate the leading derivatives in the expression for $\hat{\cal G}_4$ in Eq.\ (\ref{expandedd3calF}).  The resulting expression for $\hat{G}(v,Q^2)$ is rather lengthy, but the leading terms depend directly on $F_2^p$ and $\partial F_2^p/\partial \ln Q^2$ without further integration.  This procedure reduces the numerical sensitivity of the results to the higher  derivatives  of $F_2^p$ that appear in Eq.\ (\ref{expandedd3calF}).  The results for $G(x,Q^2)$ obtained using these two methods agree.

In the present numerical work, we used 
\be
\alpha_s(Q^2)
\equiv\frac{12 \pi}{25\ln(Q^2/\Lambda^2)},
\label{alpha}
\ee
for four active flavors, with $\Lambda=0.153$ GeV adjusted to give  $\alpha_s(M_Z^2)=0.118$.

Returning to the variable $x$, we write our final analytic answer, for $0<x\la x_P$, as
\ba
G(x,Q^2)=-\frac{1}{\omega}\int^x{dz\over z} \left({z\over x}\right)^k \sin\left(\omega\ln({z\over x})\right) {\cal G}_4(z,Q^2)&&\nonumber\\
=-0.459 - 0.143\ln(Q^2)
  -0.0155\ln^2(Q^2)\,+\hphantom{xxxxn}&&\nonumber\\
\left[0.231\!+\! 0.00971\ln(Q^2)  - 
     0.0147\ln^2(Q^2)\right]\ln(1/x)\,\,\,+&&\nonumber\\ 
\left[0.0836\! +\! 0.06328\ln(Q^2)\! +\! 
     0.0112\ln^2(Q^2)\right]\ln^2(1/x).\hphantom{i}&&\label{finalanswer}
\ea
This is a simple quadratic polynomial in $\ln(1/x)$, with quadratic polynomial coefficients in $\ln(Q^2)$. 

To make a rough   estimate of  the uncertainty in $G(x,Q^2)$ caused by the experimental uncertainties in the input function $F_2^p(x,Q^2)$, we integrate \eq{finalAPoverx} twice by parts with respect to $\ln(1/x)$ and drop terms that are suppressed by powers of $\ln(1/x)$ to get the zeroth order approximation 
\be
\tilde G(x,Q^2)\approx \frac {27}{40}\frac{4\pi}{\alpha_s}\frac{\partial  F_2^p(x,Q^2)}{\partial \ln (Q^2)}. \label{G0}
\ee
Numerical studies show that ${\cal F}_2^p(x,Q^2)$ is dominated at small $x$ by ${\partial F}_2^p(x,Q^2)/\partial \ln (Q^2)$. 
 Figure~2 of Ref.~\cite{bbt} shows the $\pm 1\sigma$ error bands of $\partial  F_2^p(x,Q^2)/\partial \ln (Q^2)$  due to the parameter errors in \eq{parameters}, including their correlations, at $x=0.00008$, as a function of $Q^2$.  For $Q^2$ in the range 5---200 GeV$^2$, it was found \cite{bbt} that  $\Delta[\partial F_2^p(x,Q^2)/\partial \ln (Q^2))]/[\partial F_2^p(x,Q^2)/\partial \ln (Q^2)]\approx 0.015-0.020$, leading to the rough estimate 
\be
\left.\Delta G(x,Q^2)/G(x,Q^2)\right |_{x \sim 0.0001}\approx 2\%,\label{errorG}
\ee
for the statistical error associated with the uncertainties in the fit to $F_ 2^p(x,Q^2)$  \cite{bbt} in \eq{bbtfit}. 

\begin{figure}[h,t,b] %Fig.3
\begin{center}
\mbox{\epsfig{file=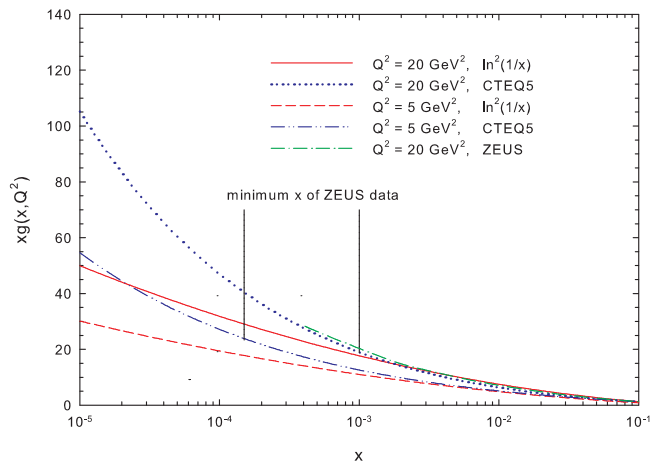,width=3.4in%
,bbllx=0pt,bblly=0pt,bburx=420pt,bbury=305pt,clip=%
}}
\end{center}
\caption[]{LO gluon distribution functions $G=xg(x,Q^2)$ for virtualities $Q^2= 5$ and 20 GeV$^2$. The curves labeled $\ln^2(1/x)$ are from our LO Froissart bound fit of \eq{finalanswer}.  The curves labeled CTEQ5 are LO CTEQ5 curves \cite{olness}. The vertical lines indicate the minimum $x$ values that the ZEUS \cite{ZEUS} collaboration achieved, for the  $Q^2$ in the plot.
\label{fig:gluon&CTEQ}}
\end{figure}

We emphasize that  only  {\em experimental $F_2^p$ data } were needed to obtain the numerical results in \eq{finalanswer} for  $x\lesssim x_P$. The 6-parameter fit of Ref. \cite{bbt} which we used  was constructed using all of the available ZEUS data  \cite{ZEUS}, with  $x_P\ge x\ga 10^{-3}$ to $10^{-4}$ depending in the virtuality, and $0.11\le Q^2 \le 1200$ GeV$^2$.  The fit was excellent, with a $\chi^2/{\rm d.o.f.}=1.09$ for 169 degrees of freedom.  The use of the low-$Q^2$ data on $F_2^p(x,Q^2)$, a virtual photon-proton scattering cross section, is important in determining the global fit.  However, we use Eq.\ (\ref{vanswer}) or Eq.\ (\ref{finalanswer}) to determine $G(x,Q^2)$ only in the perturbative region $Q^2\gtrsim 2$ GeV$^2$.

We show our LO results from \eq{finalanswer} for $G(x,Q^2)$ for $Q^2=5$ and 20 GeV$^2$ in Fig. \ref{fig:gluon&CTEQ}, in which we compare them to ZEUS \cite{ZEUS_LO} and CTEQ5 LO  power law gluon distributions \cite{olness} calculated at the same virtualities. Inspection of Fig. \ref{fig:gluon&CTEQ} shows  rough agreement for $x$-values  $\ga 10^{-3}$, where there are experimental data \cite{ZEUS} but  differ markedly for  smaller $x$, with our extrapolated fits  being {\em  much smaller} than the extrapolated CTEQ5 (and other) power law fits. 

We recall that the functions $\hat{\cal G}_4(v,Q^2)$ and ${\cal G}_4(x,Q^2)$ on the right hand sides of the analytical expressions for $G(x,Q^2)$ in Eqs.\ \ref{vanswer}  and \ref{finalanswer} depend  only on numerical values of $F_2^p(x,Q^2)$ and $\alpha_s(Q^2)$, which are in turn determined by experiment.  Thus, if two global fits to $F_2^p(x,Q^2)$---using {\em different} parametrizations---both reproduce the  experimental data in a given region of $x$ and $Q^2$ and treat $\alpha_s(Q^2)$ and the number of active flavors equivalently,  they should give the same values for the gluon distribution in that region.  This is roughly true here for the CTEQ5, ZEUS and Froissart-type fits for $x\ga 10^{-3}$, as seen in Fig.\ 1. The striking disagreements occur in  the small-$x$ region, where there are  no longer any  data for $F_2^p(x,Q^2)$ \cite{ZEUS}, and involve large extrapolations of the forms determined where data exist.

%%%%%%%%%%%%%%%%%%

{\em Conclusions.}\ \ We have shown here that, for a fixed number of active quarks, all that is actually necessary to calculate LO gluon distributions in a region of $x$ and $Q^2$ is knowledge of $F_2^p(x,Q^2)$ and $\alpha_s(Q^2)$ in that region. The integrals in \eq{vanswer} and \eq{finalanswer} give the {\em exact} LO expression for $G(x,Q^2)$ in terms of a function determined by $F_2^p(x,Q^2)$. We  used the  complete LO splitting functions $K^{(1)}_{qg}(z)$ and $K^{(1)}_{qq}(z)$, so no approximation is made at the LO level.  Further, we did  not need to use the gluon evolution equation in the derivation of our differential equation for $G(x,Q^2)$ in LO, but only the evolution equation for $F_2^p(x,Q^2)$.

Using the accurate Froissart-bound type parametrization of $F_2^p(x,Q^2)$ \cite{bbt} in \eq{finalanswer}, we obtained  an analytic solution for $G(x,Q^2)$ in the region in which there are experimental data, with $\la 2\%$ uncertainty due to fitting parameter errors of $F_2^p$.   There should be {\em no} difference between our results and other LO solutions for $x_P\ga x\ga 10^{-3}$ to $10^{-4}$ to the extent that they all reproduce the $F_2^p(x,Q^2)$ data  equally well in this $x$ region.  Presumably, the differences seen in Fig. \ref{fig:gluon&CTEQ} are due to CTEQ (and ZEUS) having used detailed model-dependent  functional forms for the quark and gluon distributions in conjunction with the DGLAP evolution in their analysis,  a procedure that involves extra assumptions, as well as possible differences in the treatment of  the number of active quarks.   Further, the CTEQ  group \cite{olness} used additional data sets.

 One of course wants information on the quark distribution functions as well as the gluon distribution for applications to high-energy particle processes, and the use of the usual DGLAP equations allows the use of data other than those on $F_2^p(x,Q^2)$, and the separate extraction of all these parton distribution functions.  We believe, however, that the present technique is simple and interesting, and gives a useful check on other approaches at LO. 

We will present our generalization of this technique to include the effects of quark masses elsewhere.  Again, we stress that {\em no} knowledge of individual quark distributions is needed.  We also will use this technique to explore the singlet quark distribution, using the gluon evolution equation. We also intend to  explore the extent to which this technique can be expanded to obtain NLO gluon distribution functions. 

{\em Acknowledgements:} M.M.B. and L.D. thank the Aspen Center for Physics for its hospitality during the time parts of this work were done.   D.W.M. receives support from DOE Grant No. DE-FG02-04ER41308.

%%%%
\end{document}